\documentclass[a4paper, conference]{IEEEtran}
\usepackage{graphicx}
\usepackage{balance}
\usepackage{nopageno}
\usepackage{cite}
\usepackage{amsthm}
\usepackage{amssymb}
\usepackage{amsmath}
\usepackage{amsfonts}
\usepackage{amsmath}
\usepackage{graphicx}
\usepackage{mathtools}
\usepackage{multirow}
\usepackage{epstopdf}
\usepackage{subfigure}
\usepackage{array}
\usepackage{csquotes}
\usepackage{caption}
\graphicspath{{Figs/}}
\usepackage{xcolor}

\newcolumntype{C}[1]{>{\centering\let\newline\\\arraybackslash\hspace{0pt}}m{#1}}

\begin{document}	
	
\title{A Comprehensive Analysis of Secondary Coexistence in a Real-World CBRS Deployment}

\author{ 
\IEEEauthorblockN{
Armed Tusha\IEEEauthorrefmark{1},
Seda Tusha\IEEEauthorrefmark{1},
Hossein Nasiri\IEEEauthorrefmark{1},
Muhammad I. Rochman\IEEEauthorrefmark{2},
Patrick McGuire\IEEEauthorrefmark{3},
Monisha Ghosh\IEEEauthorrefmark{1}\\
\IEEEauthorblockA{
\IEEEauthorrefmark{1}Department of Electrical and Electronics Engineering, University of Notre Dame, South Bend, IN, USA\\}
\IEEEauthorblockA{
\IEEEauthorrefmark{2} Department of Computer Science, University of Chicago, Chicago, IL, USA\\}
\IEEEauthorblockA{
\IEEEauthorrefmark{3}South Bend Community School Corporation, South Bend, IN, USA\\}
\IEEEauthorblockA{
Email: \{atusha, stusha, hnasiri2, mghosh3\}@nd.edu, muhiqbalcr@uchicago.edu, pmcguire@southbendin.gov}}
} 
\maketitle

\begin{abstract}
The Federal Communications Commission (FCC) in the U.S. has made the Citizens Broadband Radio Service (CBRS) band (3.55 - 3.7 GHz) available for commercial wireless usage under a shared approach using a three-tier hierarchical architecture, where the federal incumbent is the highest priority Tier 1 user, Priority Access License (PAL) holders, who have paid for licenses, are Tier 2 users and Tier 3 users operate under General Authorized Access (GAA), without license fees or protection from higher priority users.  The Spectrum Access System (SAS) ensures that higher priority users are protected from interference from lower priority users. However, the lowest priority GAA users are not given any protection from each other by the SAS and are expected to not cause any harmful interference to Tier 1 and Tier 2 users. As the deployments of GAA devices grow, the potential for secondary interference between GAA users increases, especially since the SAS architecture does not allow dynamic channel switching when faced with interference. However, there is limited academic research evaluating the real-world system performance of GAA deployments and secondary coexistence. In this paper, we present a first-of-its-kind extensive measurement campaign of a commercial CBRS network deployed in the city of South Bend, IN, 
that quantifies both co-channel interference (CCI) and adjacent channel interference  (ACI) caused by competing GAA devices and C-band 5G, respectively. We (i) identify a particular CCI scenario and improve performance by changing the frequency allocation based on our study of other allocations in the vicinity and (ii) quantify ACI from 5G in C-band (3.7 GHz) on CBRS throughput. We conclude that (i) CCI and ACI for GAA users is not handled well by the SAS, (ii) proper frequency allocation for GAA requires additional analysis of interference from other GAA users followed by dynamical channel selection, and (iii) utilization of immediate adjacent channels by high power 5G deployments limits the performance of CBRS. 
\end{abstract}


\IEEEpeerreviewmaketitle

\section{Introduction}\label{introduction}
\subsection{Overview of CBRS Band}

\begin{figure}[t]
	\centering	\includegraphics[scale=0.26]{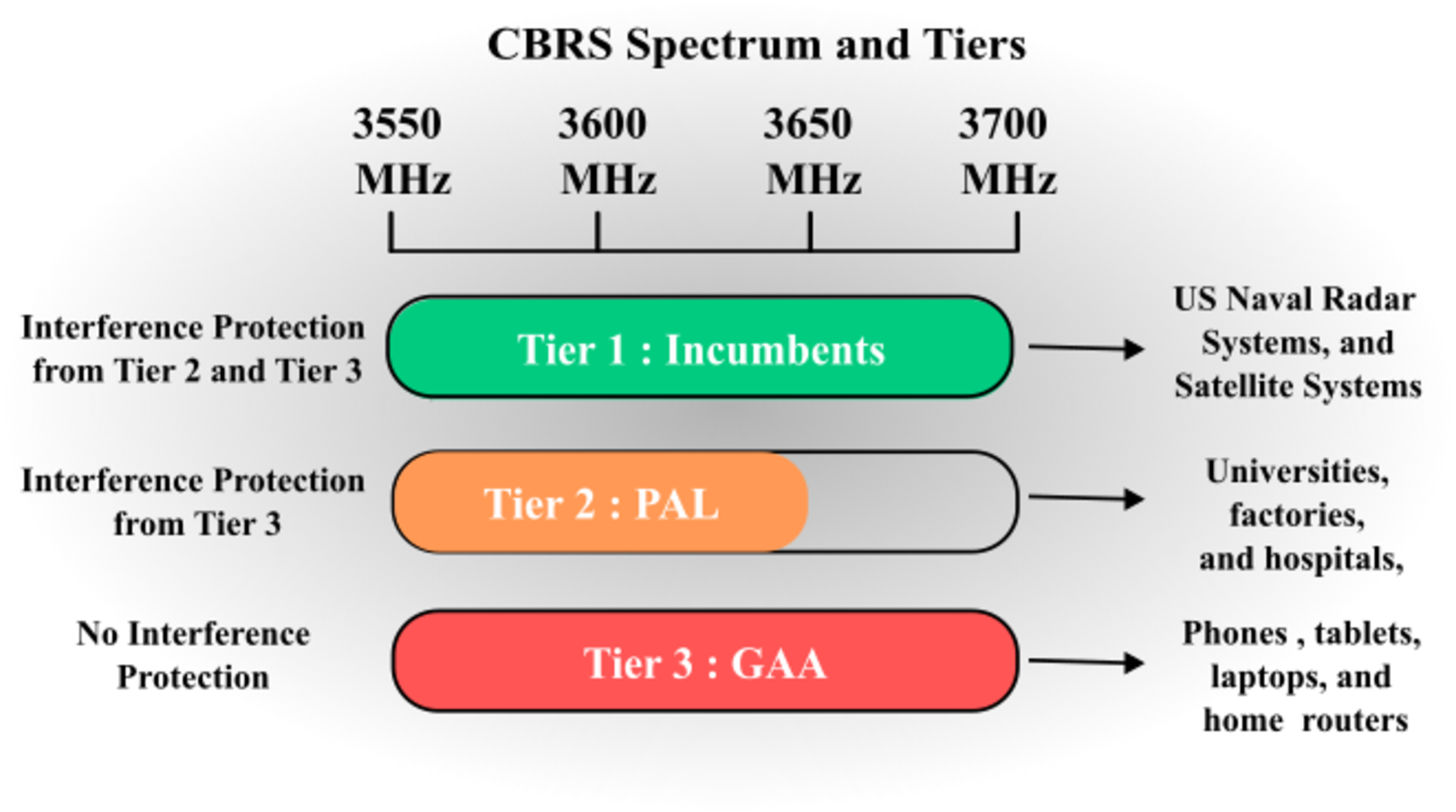}
	\caption{Three-tier hierarchical architecture of CBRS band.}
	\label{Fig:SAS}
\end{figure}

The demand for spectrum is  growing rapidly with the advancement of wireless technologies that deliver new services and applications \cite{shah20185g}. 
Increasingly, new spectrum will be shared with federal services. In April 2015, the U.S. Federal Communications Commission (FCC) authorized the Citizens Broadband Radio Services (CBRS) band, 3.55 - 3.7 GHz, for shared use by commercial wireless providers, while protecting the incumbent federal user, primarily Navy radar. As shown in Fig.~\ref{Fig:SAS}, CBRS users are grouped into three different tiers based on their spectrum access priorities \cite{ghosh2019competition}. Incumbent users are designated as Tier 1 \cite{exoplanetwebsite} and must be kept interference-free from both Tier 2 and Tier 3, known as Priority Access License (PAL) and Generalized Authorized Access (GAA), respectively \cite{agarwal2022survey}. The transmit power of Tier 2 and Tier 3 users is capped at 30 dBm/10 MHz for indoor use and 47 dBm/10 MHz for outdoor use. Tier 2 (PAL) and Tier 3 (GAA) users are primarily commercial wireless service providers deploying public and private wireless networks using 4G, 5G and proprietary technologies \cite{ross2019annual}. 

PAL operation is limited to the channels between 3550 MHz - 3650 MHz while GAA users have access to the entire 150 MHz band, but can access only those channels that are not being occupied by Tier 1 and Tier 2 users for a given frequency, time and area \cite{dogan2023evaluating}. The Spectrum Access System (SAS) controls access by both PAL and GAA users, both of which can transmit only after SAS authorizes channels for their use while protecting higher priority users: this is not a dynamic process since the SAS performs aggregate interference calculations every time channels are assigned to new users. However, as CBRS use grows, this can pose a problem since the interference environment faced by GAA devices can change rapidly necessitating quick changes in the frequency of operation.

Several companies have deployed private 4G Long Term Evolution (4G LTE) and 5G networks in public venues such as manufacturing plants, industrial internet-of-things, smart homes, stadiums, universities and other use cases, using CBRS \cite{paolini2019cbrs}\cite{kim2015design}. Mobile network operators such as Verizon also utilize the CBRS spectrum in addition to their traditional licensed channels, using Carrier Aggregation (CA) and Dual Connectivity (DC), to increase the overall throughput. Competing GAA users in the same area thus experience co-channel interference (CCI) whether they are from the same deployment or from different deployments. 

Furthermore, the 3.45 - 3.55 GHz and 3.7 - 3.98 GHz bands are immediately adjacent to CBRS and are allocated for exclusively licensed cellular networks \cite{maeng2023sdr}, with permitted power spectral densities (PSDs) of 62 dBm/MHz for urban areas and 65 dBm/MHz for rural areas. These power levels are much higher than those of Citizens Broadband Radio Services Devices (CBSDs). C-band (3.7 - 3.98 GHz) services are being extensively deployed across various regions in the U.S. and 3.45 GHz services are in the early stages of deployment\cite{verizon}. This leads to potential adjacent channel interference (ACI) from 3.45 GHz and C-band to CBSDs operating at the edges of the CBRS band. In \cite{rochman2023measurement}, the authors performed detailed measurements and analyses of a real-world C-band deployment adjacent to an indoor CBRS deployment, where it was shown that the ACI level decreased by introducing a 20 MHz guard band between C-band and CBRS band. The study reported in this paper focuses on the analysis of secondary coexistence between GAA users in a real-world, outdoor CBRS deployment which faces CCI and ACI from its own network as well as an operator deployed network in CBRS and C-band.  

%
%
\subsection{Motivation \& Main Contributions}
Given the above discussion, the aim of this study is to evaluate the performance of a real-world, outdoor CBRS deployment, with a focus on secondary coexistence between GAA users and adjacent channel coexistence with C-band 5G deployments. The main contributions of this paper are as follows.

\begin{itemize}
    \item Extensive outdoor measurements of a commercial CBRS network in South Bend, IN, deployed by the local school district to provide affordable wireless broadband connectivity for students and their families in the area. The deployment consists of four base-stations (BSs), each with multiple CBSDs serving different sectors on different channels. 

    \item Generating coverage heatmaps of throughput for each CBSD to evaluate the effect of different system parameters including height, frequency of operation, foliage and interference. We observed outdoor throughput of up to 140 Mbps, in the absence of interference, but CCI, ACI, height of transmitter and foliage were significant factors in reducing performance in many areas.

    \item We faced two interference scenarios in the deployment: (i) interference within the BS and between BSs due to frequency reuse by CBSDs, and (ii) interference caused by the utilization of both CBRS and C-Band by Verizon in the same region. Based on our study of frequency allocations and measurements of signal strengths in the vicinity of the deployment, we proposed a new frequency allocation to avoid CCI between two CBSDs. This resulted in an increase of 1 - 3 dBm in reference signal received power (RSRP) and 1 - 4 dB in reference signal received quality (RSRQ), and consequently, usage of higher-order modulation and coding scheme (MCS) thus demonstrating the importance of careful frequency allocations based on real-time interference measurements.

  \item  Compared to other CBRS channels, we observed a decrease of 12 Mbps in median downlink (DL) throughput of CBSDs operating on 3690 MHz due to 5G deployments in the adjacent C-band channel, 3700 - 3760 MHz, by Verizon.

  \item Along with 5G in C-band, Verizon has also deployed CBSDs throughout the CBRS band and LTE using CA (LTE-CA) utilizes up to five 20 MHz CBRS channels delivering throughput that exceeds 5G throughput over C-band: this indicates that CBRS use by mobile operators will impact performance of smaller, private networks such as those deployed by the South Bend school district.
\end{itemize}

\begin{figure}
	\centering
	\includegraphics[scale=0.24]{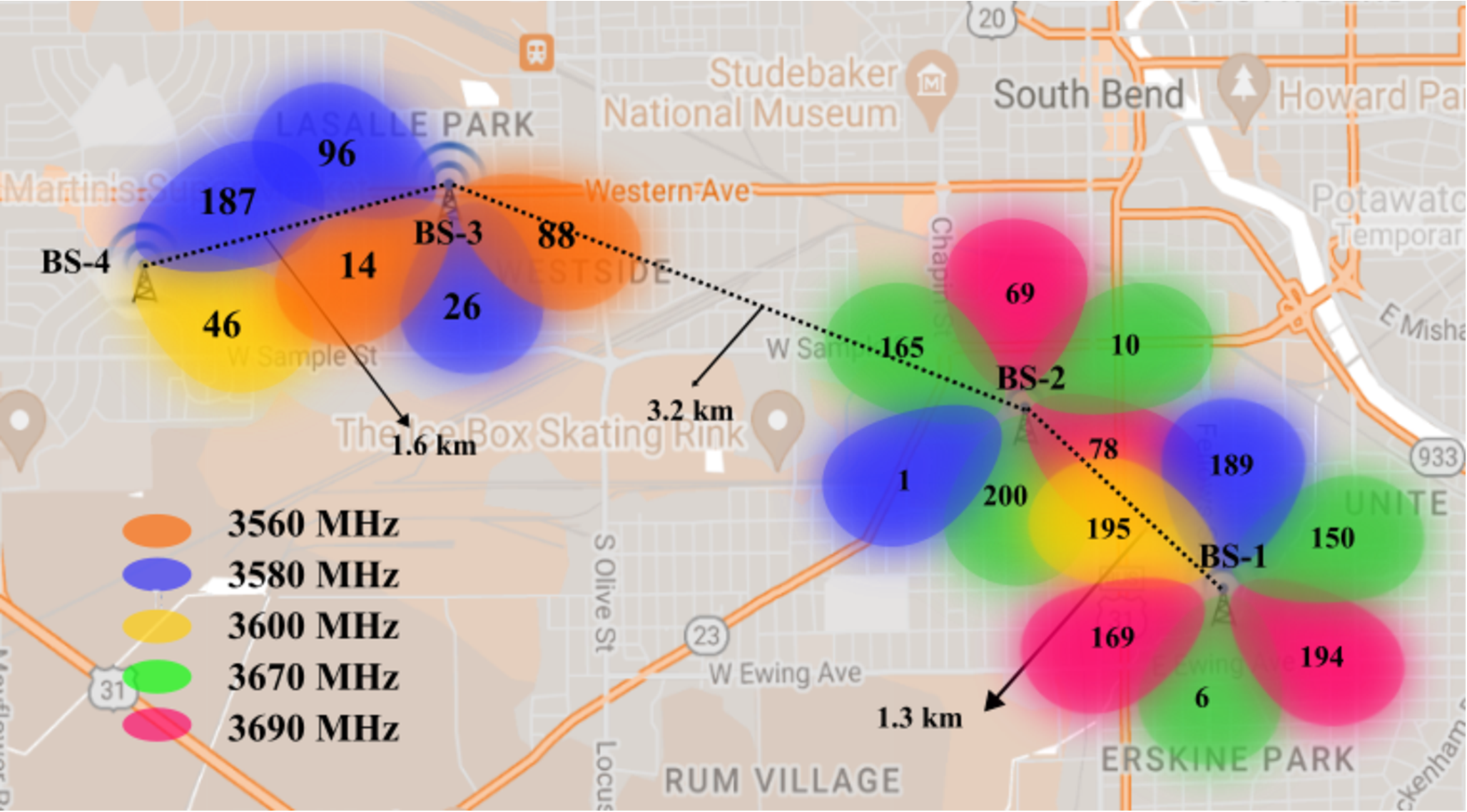}
	\caption{CBRS deployment by the South Bend school district.}
	\label{Fig:CBRSdeployment}
\end{figure}

\begin{table}
	\caption{Height, PCI, and the center frequencies of the channel allocations for BSs.}
	\centering
	\small	
 \renewcommand{\arraystretch}{1.1}
	\begin{tabular}{|C{1.2cm}|C{1cm}|C{0.65cm}|C{0.65cm}|C{0.65cm}|C{0.65cm}|C{0.65cm}|} 
 \hline
\multirow{2}{*}{\textbf{Region}} & \multirow{2}{*}{\textbf{Height}} & \multicolumn{5}{c|}{\textbf{Channel Allocations (Center freq.)}}\\ 
\cline{3-7}
& \textbf{(m)} & 3560 MHz  & 3580 MHz & 3600 MHz & 3670 MHz & 3690 MHz\\
  \hline \hline BS-1 (6 PCIs) & 45 & & 189 & 195 & 6, 150 & 169, 194\\
     \hline
  BS-2 (6 PCIs) & 55 & & 1 & & 10, 200, 165& 78, 69\\
\hline
  BS-3 (4 PCIs) & 33 & 14, 88 &  96, 26 & & &\\
\hline
  BS-4 (2 PCIs) & 13 & & 187 & 46& &\\
\hline
	\end{tabular}
	\label{tab:Freq_vs_PCI} 
\end{table}

\begin{table}[t]
	\caption{Measurement tools and devices.}
	\centering
	\small	\renewcommand{\arraystretch}{1.1}
	\begin{tabular}{|C{1.8 cm}|C{3cm}|C{2.6cm}|} 
 \hline
 \textbf{App./Tool} & \textbf{Features}  & \textbf{Devices} \\
  \hline \hline  SigCap &  Operator,PCI, EARFCN, Band, Frequency, Altitude, Longitude, Latitude, RSRP, RSRQ, RSSI & 1 $\times$ Google P5, \newline 1 $\times$ Google P6, \newline 1 $\times$ Samsung S21 \\
     \hline
  QualiPoc & Operator, PCI, Band, Altitude, Longitude, Latitude, RSRP, RSRQ, CQI, RSSI, DL/UL Throughput, RB per subframe & 2 $\times$ Samsung S22+ \\
\hline
  PRiSM & PCI, EARFCN, Frequency, Altitude, Longitude, Latitude, RSRP, RSRQ, RSSI & 1 $\times$ Google P5 \\
\hline
\end{tabular}
\label{tab:FeatureDevices} 
\end{table}

%
%
\section{DEPLOYMENT, TOOLS \& METHODOLOGY  }\label{CBRSdeployment}





%
%

\subsection{Deployment}
An extensive measurement campaign was conducted over a wide area of approximately 12 km$^2$  where the South Bend school district has deployed CBRS, as shown in Fig. \ref{Fig:CBRSdeployment}.
Four BSs, i.e., James Whitcomb Riley High School (BS-1), Hayes Tower (BS-2), West Tower (BS-3), and Navarre Middle School (BS-4) have been deployed on school buildings and towers, allowing South Bend schools to launch its own private LTE network to serve students and families. To sustain a high throughput and enhance system capacity, each BS has multiple CBSDs, each operating on a separate sector at the maximum permitted power of 47 dBm/10 MHz. Each CBSD is identified by its Physical Channel Identity (PCI) and operates over a single 20 MHz wide channel. Table \ref{tab:Freq_vs_PCI} provides the details about these BSs including their height, PCI, and the center frequencies of the channel allocations. Since there are only 7 non-overlapping 20 MHz channels in the CBRS band and the deployment has 18 PCIs, it is clear that CBSDs will reuse channels. Thus, channel reuse is introduced at each BS via sectorization at the expense of potential CCI. It should be noted that the Google SAS shows that all 15 channels (10 MHz each) are available for GAA use by this deployment.

\textit{BS-1} is deployed on the roof of James Whitcomb Riley High School at a height of 40 m. It uses four 20 MHz channels and six sectors with the PCIs as shown in Table \ref{tab:Freq_vs_PCI}.

\textit{BS-2} is mounted on a tower with a height of 55 m, and uses three channels across six PCIs, three of which operate on the same frequency (3670 MHz). The distance between BS-1 and BS-2 is about 1.3 km. 

 \textit{BS-3} is located on a tower at a lower height (33 m), compared to BS-1 and BS-2, and two channels, i.e., 3560 MHz and 3580 MHz, are used to serve four sectors with PCIs as shown in Table \ref{tab:Freq_vs_PCI}.
 
 \textit{BS-4} is placed on the roof of Navarre Middle School at the lowest height (13 m), and uses 3580 MHz and 3600 MHz to serve PCIs 187 and 46, respectively. The sector with PCI 187 is directed towards BS-3 with PCIs 96 and 26 operating on the same channel, representing potential CCI.

 \subsection{Measurement Tools}
Smartphones were used as user equipment (UEs) to capture detailed signal information, using tools such as SigCap, QualiPoc, and PRiSM as shown in Table \ref{tab:FeatureDevices}.

\textit{SigCap} is an Android application which collects wireless signal parameters (cellular and Wi-Fi) by using APIs without requiring root access \cite{9261954}. It allows extraction of detailed signal parameters such as Received Signal Strength Indicator (RSSI), RSRP, RSRQ, channel band and frequency for 4G, 5G, and Wi-Fi technologies every 5 seconds, along with location and time-stamps from the GPS receiver on the device.

\textit{QualiPoc} is a commercial measurement application developed by Rohde \& Schwarz and installed on Android phones \cite{QualiPocref}. In addition to signal parameters extracted by SigCap, QualiPoc collects MCS, block error rate (BLER),  time division duplexing (TDD) configuration, channel quality indicator (CQI), and physical layer throughput. All DL throughput results discussed in this work are extracted from QualiPoc, running an iperf utility.

\textit{PRiSM} is a software-defined radio (SDR) based handheld network scanner for surveying 4G/5G networks and also operates as a spectrum analyzer from 70 MHz to 6 GHz \cite{prismref}. It easily connects to PCs, tablets, and smartphones to monitor the frequency of interest. Unlike the above two tools, PRiSM does not require a SIM card to extract network information and uses the smartphone merely as a display and recording device to track channel occupancy.

\begin{figure}
	\centering
	\begin{subfigure}[The distance of measurement location from the BS-1 and nearby Verizon BS.]{\includegraphics[scale=0.20]{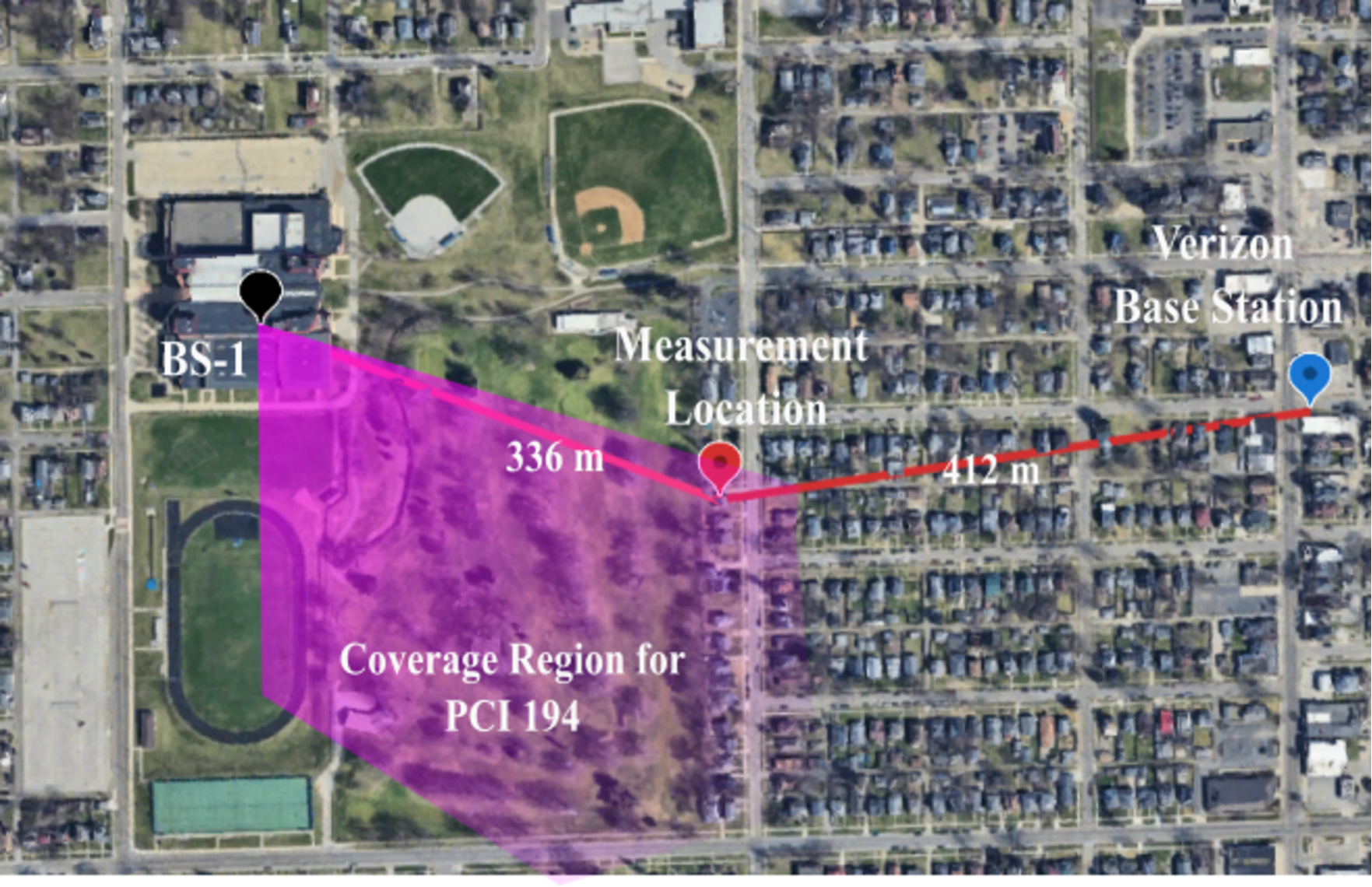}\label{Fig:loc-test3}}
	\end{subfigure} 
	\begin{subfigure}[Measurement setup.]
{\includegraphics[scale=0.20]{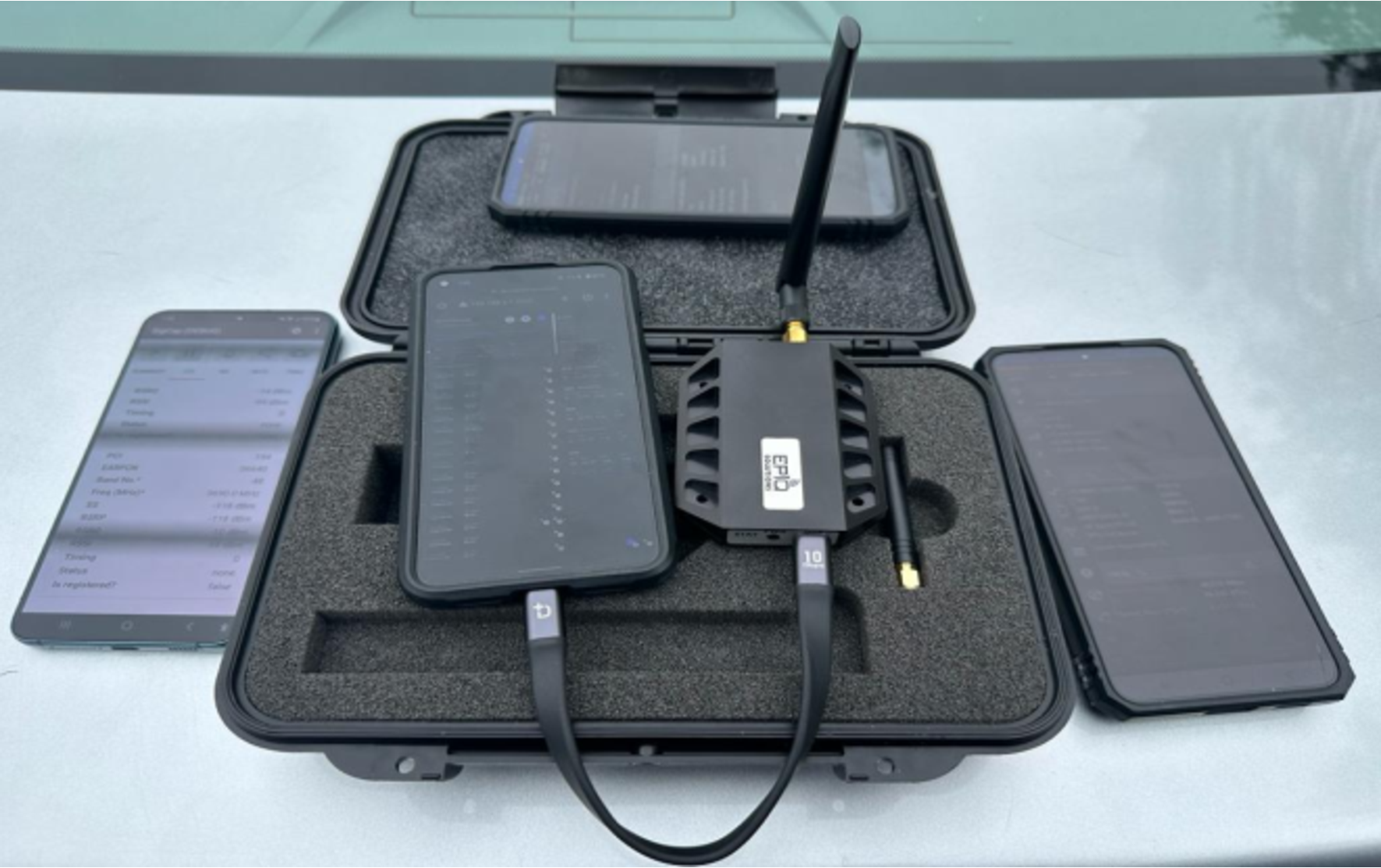}\label{Fig:Setup-test3}}
	\end{subfigure}
	\caption{Measurement environment and setup for MC-3.} 
	\label{Fig:test3}
\end{figure}

%
%
\subsection{Measurement Campaigns (MCs)}
Driving and stationary measurements campaigns (MCs) were conducted during the summer months of 2023, with dense foliage-covered trees. The UEs connected to the CBRS network using SIM cards provided by the school district. Measurements of the Verizon network used a 5G SIM with an unlimited data plan and no throttling. We grouped our experiments into three separate campaigns.

\subsubsection{MC-1} 
\begin{figure} [t]
	\centering
	\begin{subfigure}[Throughput performance.]
		{\includegraphics[width=7 cm,height= 4 cm]{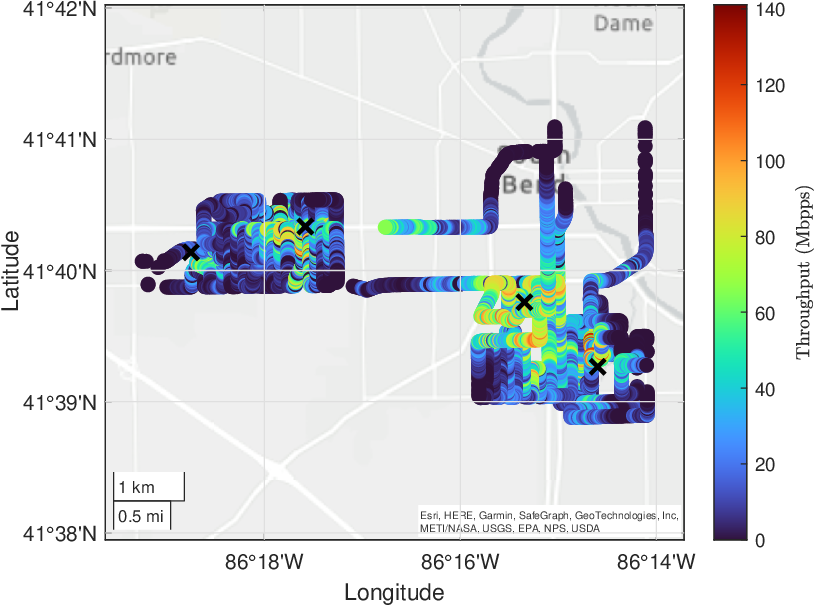}\label{Fig:all_throughput_map}}
	\end{subfigure} 
 	\begin{subfigure}[Average RBs/subframe per PCI.]
		{\includegraphics[width=7 cm,height= 4 cm]{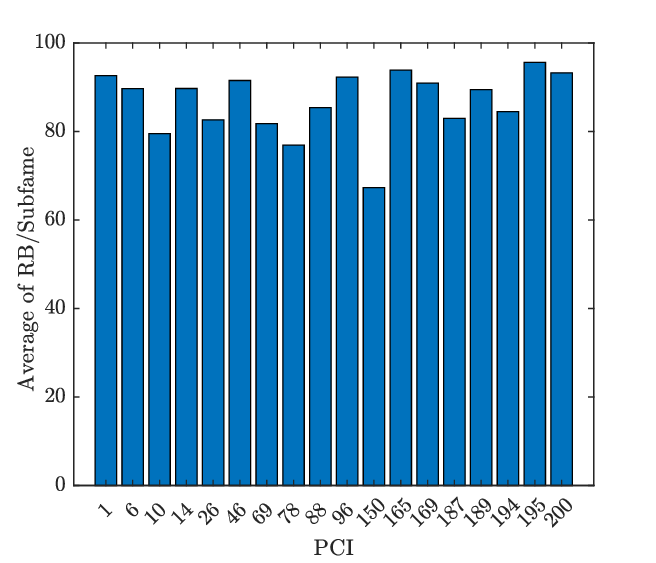}\label{Fig:ALL_PCI_RB}}
	\end{subfigure}  
	\caption{Throughput performance and average RBs/subframe for each BS.} 
	\label{Fig:All_BSs_Eval}
\end{figure}

Driving measurements were conducted around all the CBSDs as shown in Fig.~\ref{Fig:all_throughput_map}, at an average speed of 32 km/hour, over a time period of nearly 3 hours per CBSD. QualiPoc, SigCap and PRiSM were used to collect data, running on the smartphones shown in Table \ref{tab:FeatureDevices}. DL throughput measurements were recorded on two Samsung S22+ phones with QualiPoc, while the PRiSM was connected to a Google P5 and scanned all CBRS and C-Band channels in order to identify other users using these bands.

%
%
\subsubsection{MC-2} This campaign focused on BS-3 and BS-4, which are 1.6 km apart, to evaluate potential CCI in the deployment due to reuse of 3580 MHz by CBSDs in these two BSs. After identifying CCI, we worked with the network provider to change frequency assignments and evaluated the improvement when CCI was removed.

%
%
\subsubsection{MC-3} To evaluate ACI due to C-band, we conducted focused, stationary, measurements between PCI 194 deployed on 3690 MHz in BS-1 and a nearby Verizon BS operating in 3700 - 3760 MHz. Fig. \ref{Fig:loc-test3} shows the measurement location for MC-3,  and its distance from BS-1 and the Verizon BS. The experiments were conducted in two phases to assess the performance of PCI 194 on 3690 MHz under ACI caused by the usage of C-Band. CBRS and C-Band users first conducted DL transmissions at different time instants, avoiding ACI. Then, they performed simultaneous DL transmissions, leading to ACI on CBRS band. Fig. \ref{Fig:Setup-test3} shows the devices used during MC-3. PRiSM was used to continuously monitor CBRS (Band 48) and C-band (n77/n78) usage.

\begin{figure}
	\centering
	\begin{subfigure}[]
		{\includegraphics[width=7.0 cm,height= 4 cm]{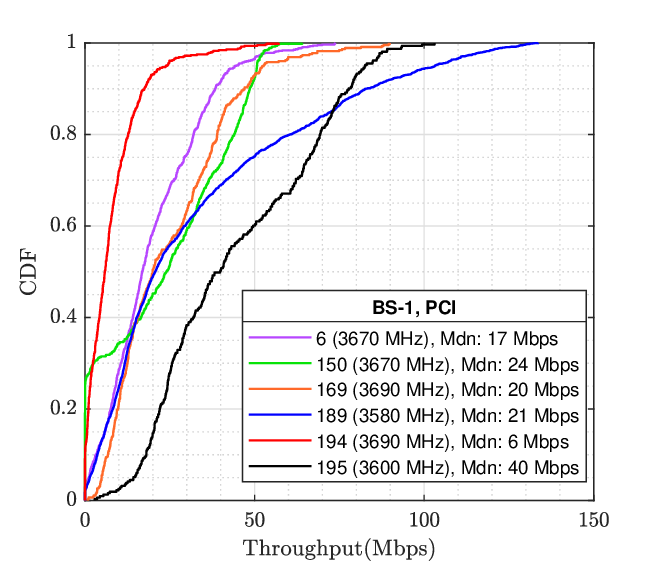}\label{Fig:riley_through}}
	\end{subfigure} 
	\begin{subfigure}[]
		{\includegraphics[width=7.0 cm,height= 4 cm]{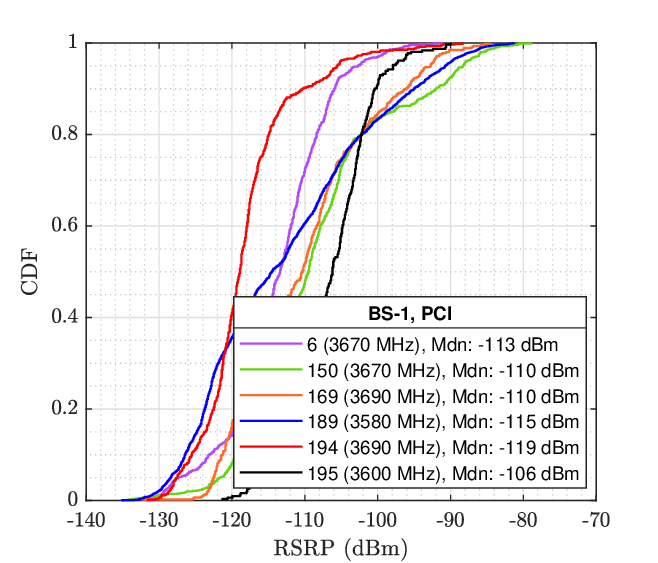}\label{Fig:riley_RSRP}}
	\end{subfigure} 
 \begin{subfigure}[] 
		{\includegraphics[width=7.0 cm,height= 4 cm]{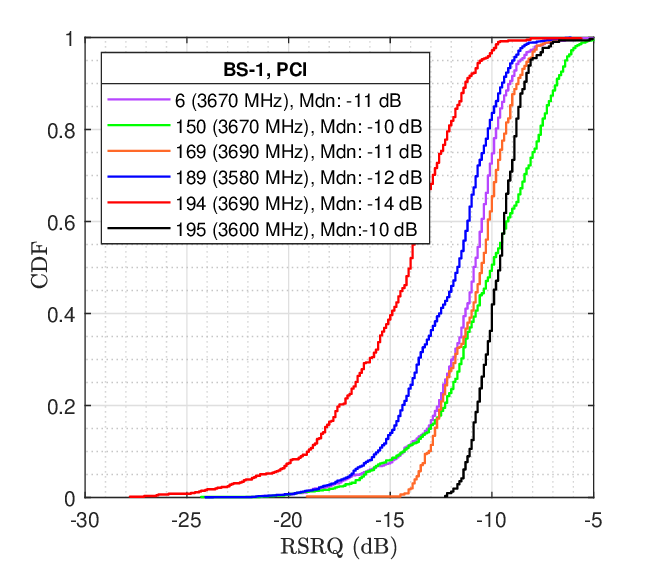}\label{Fig:riley_RSRQ}}
	\end{subfigure} 
\begin{subfigure}[] 		
         {\includegraphics[width=6.8 cm,height= 4 cm]{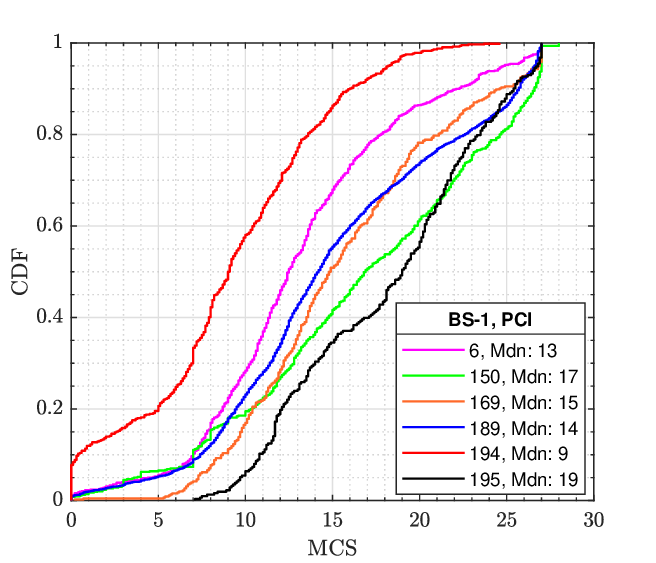}\label{Fig:riley_mcs194_195}}
	\end{subfigure}
	\caption{ CDF plots of throughput, RSRP, RSRQ and MCS for PCIs at BS-1. } 
	\label{Fig:riley_1}
\end{figure}

%
%
\section{Performance Results and Discussions}
In this section, we present statistical analyses of the measurements under different conditions. The discussion is divided into four main categories: i) the performance of a real-world CBRS deployment, ii) CCI amongst GAA users, iii) ACI from C-band to CBRS, and iv) utilization of CBRS band by mobile operators and comparison with C-band.

%
%
\subsection{Performance Evaluation}

Fig. \ref{Fig:all_throughput_map} illustrates the map of outdoor throughput obtained via driving measurements and Fig. \ref{Fig:ALL_PCI_RB} shows the distribution of RBs per subframe across all the PCIs in the deployment.  The outdoor throughput observed is in the range of 20-40 Mbps on average. The highest and lowest throughput were observed around BS-2 and BS-4 due to the height of these BSs at 55 m and 13 m, respectively. Throughput observed around BS-2 is notably higher compared to other BSs, while the lowest tower height of BS-4 leads to only a small area with high throughput.
BS-1 is at a comparable height to BS-2; however, its coverage area is notably smaller than that of BS-2 due to dense tree coverage, especially to the southeast. Since throughput is primarily a function of number of RBs allocated and MCS, we verify that the differences in measured throughput are not primarily due to RB allocation: Fig. \ref{Fig:ALL_PCI_RB} shows that the RB usage is approximately similar, with some differences that will be addressed later.

\begin{figure}
	\centering
	\includegraphics[width=8.0 cm,height= 4.5 cm]{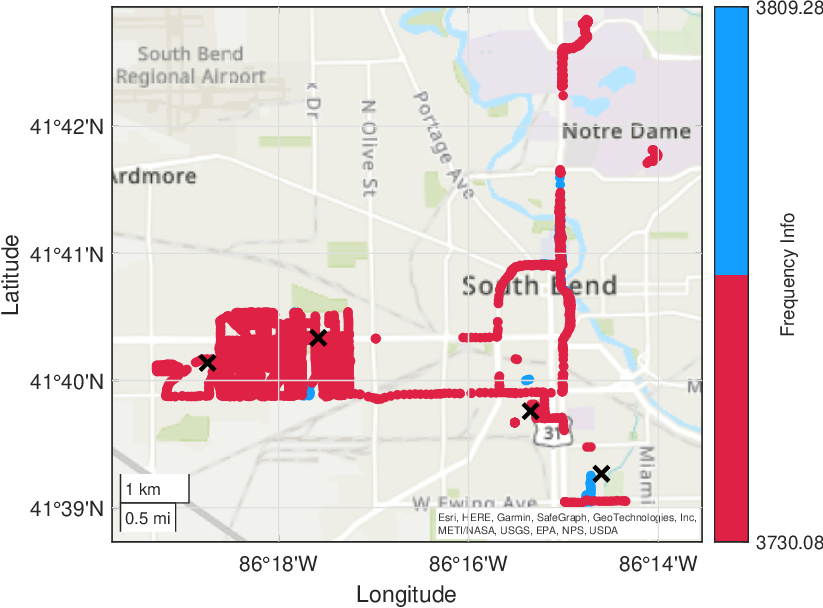}
	\caption{The usage of C-Band by Verizon in the region.}
	\label{Fig:vrz_cband_map}
\end{figure}

We analyzed the measured throughput, RSRP and RSRQ for each BS using cumulative distribution functions (CDFs). 

\textit{Performance of BS-1:} Fig. \ref{Fig:riley_1} presents the results for each PCI of BS-1. PCI 195 has the best throughput in Fig. \ref{Fig:riley_through}, almost double that of the other PCIs, as it is the only PCI from BS-1 or BS-2 operating on 3600 MHz, as seen from Table \ref{tab:Freq_vs_PCI}, and thus faces no CCI from other PCIs on the same channel. Similarly, the RSRP and RSRQ for PCI 195 outperforms the other PCIs in BS-1 as given in Figs. \ref{Fig:riley_RSRP} and \ref{Fig:riley_RSRQ}, respectively. PCIs 6 and 150 operate on 3670 MHz, while PCIs 169 and 194 operate on 3690 MHz. Although PCIs 6 and 150 showed similar median throughput performances at 17 Mbps and 24 Mbps, respectively, there is a substantial performance gap between PCI 169 and PCI 194, achieving 20 Mbps and 6 Mbps respectively. PCI 194 also exhibited the worst RSRP, RSRQ  and MCS performance as compared to the best performing PCI 195 in BS-1 as seen from \ref{Fig:riley_mcs194_195}, which explains the lower throughput. Based on our detailed analysis of signal strength measurements in the vicinity of BS-1, the reason for this is that PCI-194 experiences ACI due to the use of the immediately adjacent C-band, by a nearby Verizon BS, as shown in Figs. \ref{Fig:loc-test3} and \ref{Fig:vrz_cband_map}. The performance of PCI 194 both with and without C-band usage, will be discussed in Section III.C below.

\begin{figure}
	\centering
	\begin{subfigure}[]
		{\includegraphics[width=7.0 cm,height= 4 cm]{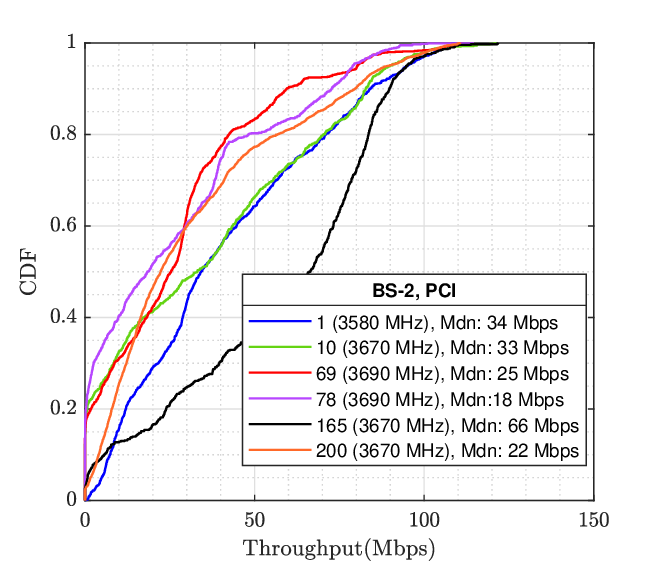}\label{Fig:hayes_through}}
	\end{subfigure} 
	\begin{subfigure}[]
		{\includegraphics[width=7.0 cm,height= 4 cm]{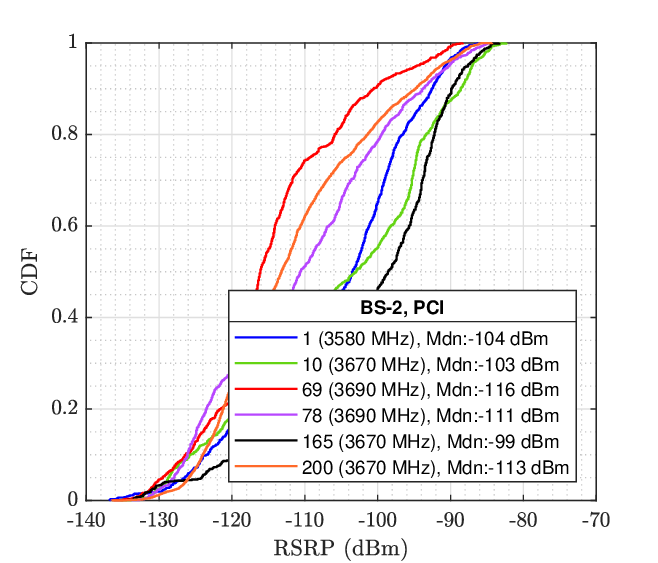}\label{Fig:hayes_RSRP}}
	\end{subfigure} 
 \begin{subfigure}[]
 {\includegraphics[width=7.0 cm,height= 4 cm]{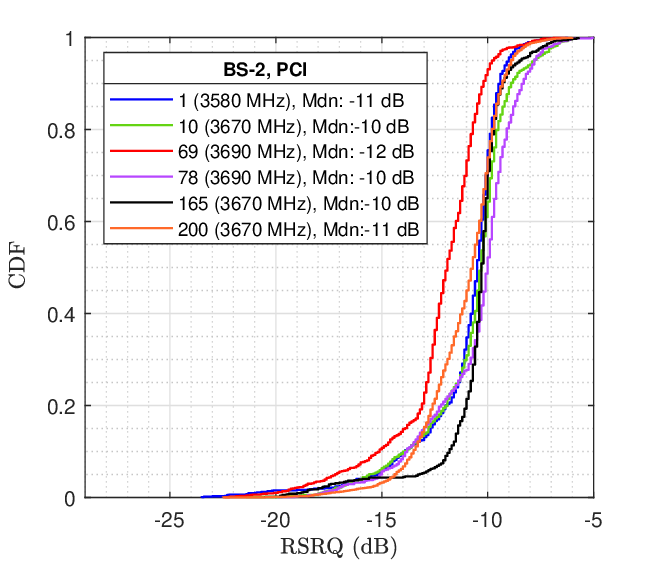}\label{Fig:hayes_RSRQ}}
	\end{subfigure} 
	\caption{ CDF plots of throughput, RSRP and RSRQ for PCIs at BS-2.} 
	\label{Fig:hayes_1}
\end{figure}

\begin{figure*}
	\centering
	\begin{subfigure}[]
		{\includegraphics[width=6.2 cm,height= 4 cm]{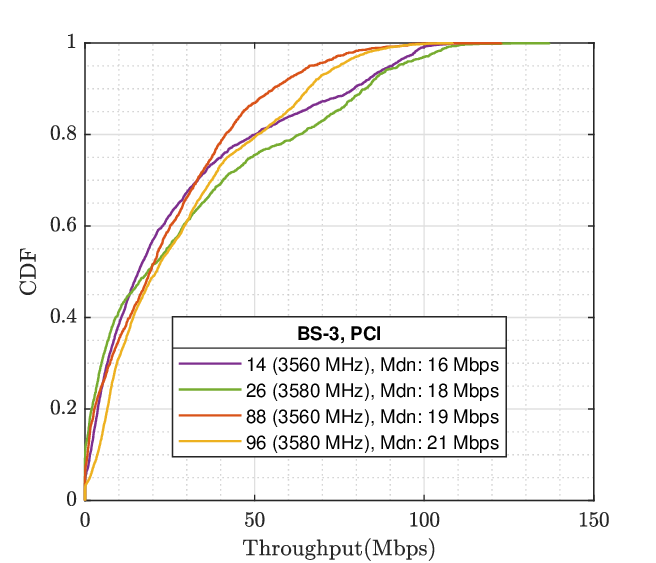}\label{Fig:west_through}}
	\end{subfigure} \hspace{-2em}
	\begin{subfigure}[]
		{\includegraphics[width=6.2 cm,height= 4 cm]{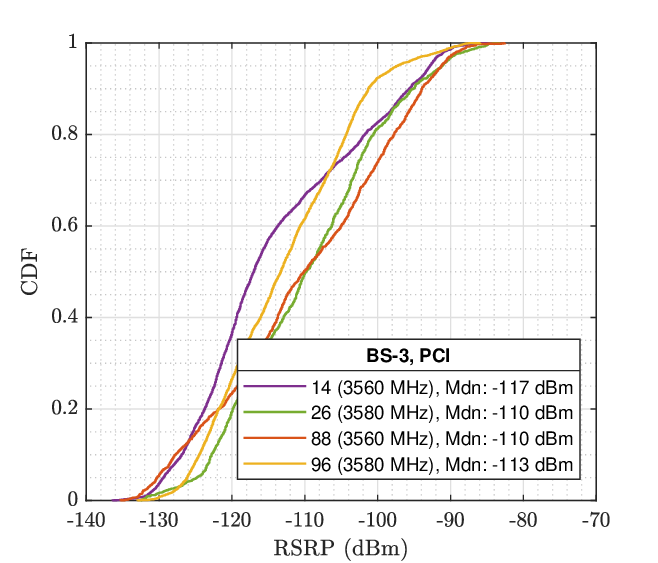}\label{Fig:west_RSRP}}
	\end{subfigure} \hspace{-2em}
 \begin{subfigure}[]
		{\includegraphics[width=6.2 cm,height= 4 cm]{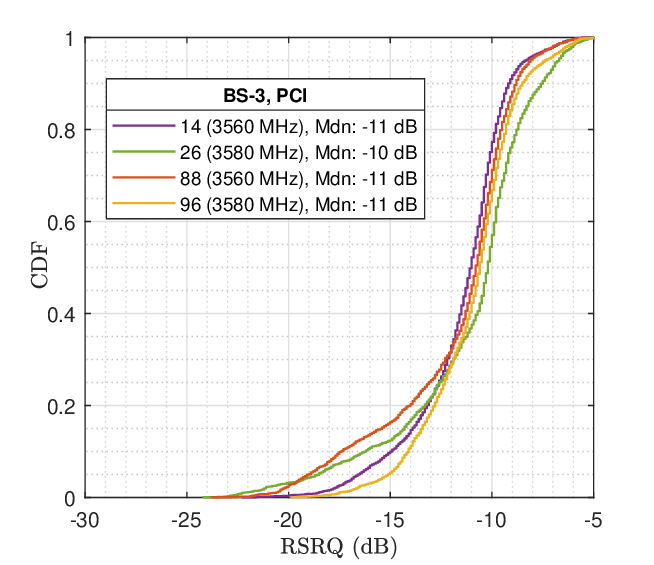}\label{Fig:west_RSRQ}}
	\end{subfigure} 
	\caption{ CDF plots of throughput, RSRP and RSRQ for PCIs at BS-3.} 
	\label{Fig:west_1}
\end{figure*}

\begin{figure*}
	\centering
	\begin{subfigure}[]
		{\includegraphics[width=6.2 cm,height= 4 cm]{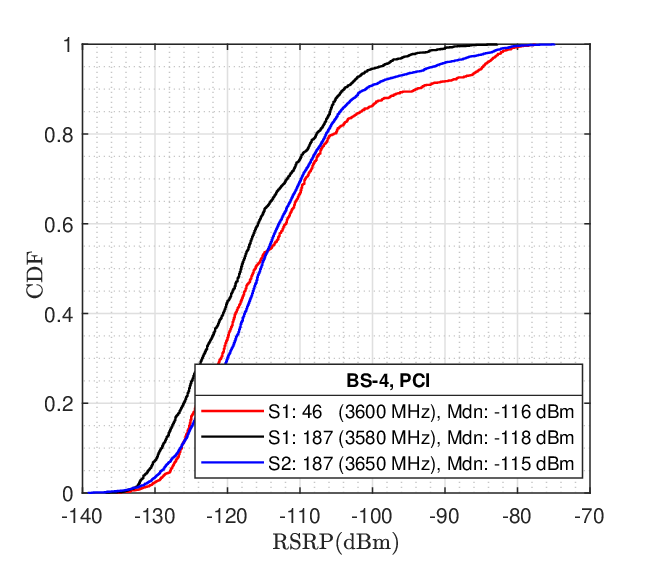}\label{Fig:Nav_RSRP}}
	\end{subfigure} \hspace{-2 em}
 \begin{subfigure}[]
		{\includegraphics[width=6.2 cm,height= 4 cm]{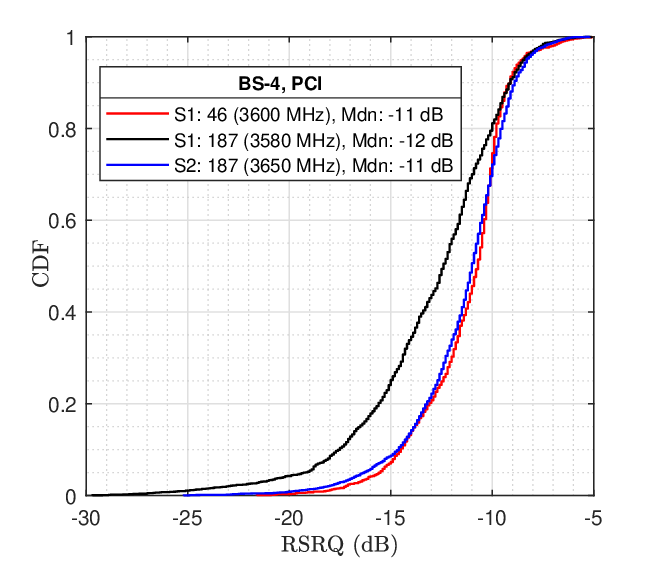}\label{Fig:Nav_RSRQ}}
	\end{subfigure} \hspace{-2 em}
 \begin{subfigure}[] 
		{\includegraphics[width=6.2 cm,height= 4 cm]{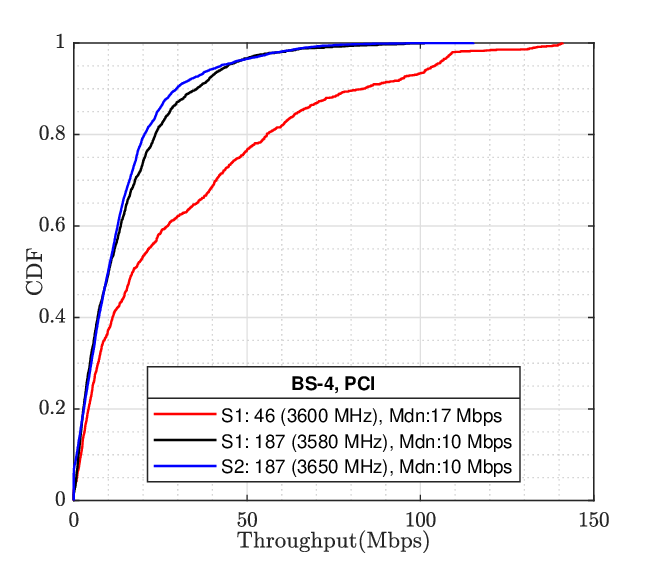}\label{Fig:Nav_through}}
	\end{subfigure} 
	\caption{ CDF plots of RSRP, RSRQ, and throughput for BS-4 i) Scenario 1 (S1): PCI 187 on 3580 MHz and PCI 46 on 3600 MHz,  and ii) Scenario 2 (S2): PCI 187 on 3650 MHz. } 
	\label{Fig:Navarre_1}
\end{figure*}

\begin{figure}
	\centering
	\includegraphics[width=7.5 cm,height= 5 cm]{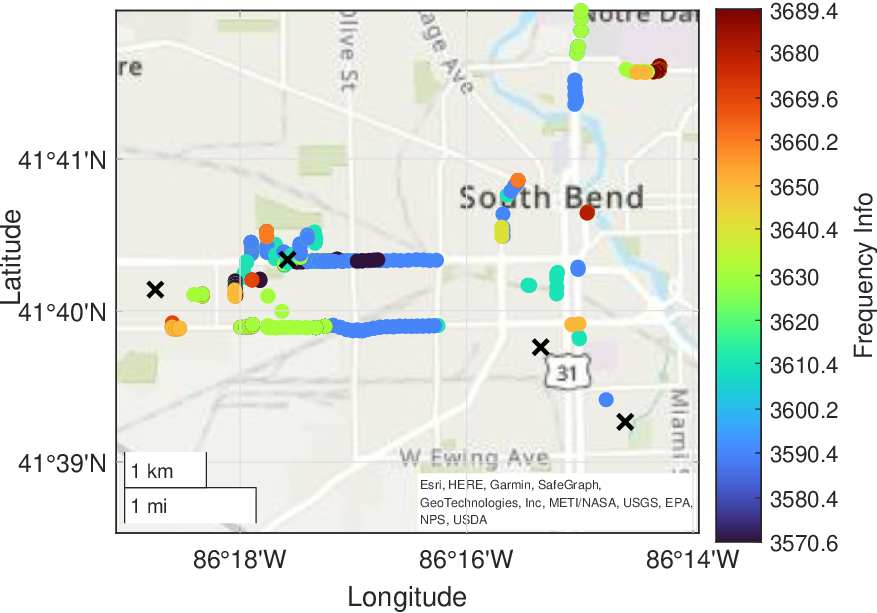}
	\caption{CBRS band usage by Verizon in the region.}
	\label{Fig:vrz_cbrs_map}
\end{figure}
\textit{Performance of BS-2:} Fig. \ref{Fig:hayes_1} shows the performance of BS-2, operating at a height similar to BS-1, but within an area where free-space propagation is more prevalent. Hence, PCI 165 on 3670 MHz achieves the highest median throughput of the CBRS deployment, reaching 66 Mbps as indicated in Fig. \ref{Fig:hayes_through}. The median throughput of PCI 165 is nearly double that observed on the other PCIs on 3670 MHz in BS-2 and BS-1, as PCI 165 is the only one directed northwest, while the rest are oriented southeast, potentially leading to CCI among them. PCI 200 has the lowest median throughput, 22 Mbps, amongst the PCIs on 3670 MHz in BS-2 due to its orientation toward a residential area with dense trees. As discussed for PCI 194 in BS-1, we observe that PCIs 69 and 78 on 3690 MHz have the worst performance in BS-2: this can be explained by ACI resulting from the usage of 3730 MHz in the vicinity of BS-2, as shown in Fig. \ref{Fig:vrz_cband_map}.  PCI 78 has the lowest median throughput of 18 Mbps on BS-2, and 7 Mbps lower than PCI 69, since its coverage overlaps with PCI 169 on 3690 MHz coming from BS-1, as seen in Fig. \ref{Fig:CBRSdeployment}.
RSRP results in Fig. \ref{Fig:hayes_RSRP} clearly exhibit the reduced impact of foliage on BS-2, where three PCIs (165, 1 and 10) have a median greater than -105 dBm. Similarly, in Fig. \ref{Fig:hayes_RSRQ}, the median RSRQ levels in BS-2 ranged from -12 dB to -10 dB, and provided better performance than BS-1. As in the throughput results, PCI 69 on 3690 MHz offered the lowest RSRQ performance due to the ACI.

\textit{Performance of BS-3:} Fig. \ref{Fig:west_1} presents the results of BS-3, which is less likely to suffer from interference since the distance of 3.2 km between BS-3 and BS-2 mitigates the presence of CCI, while the utilization of the lower edge of the CBRS spectrum (3560 MHz and 3580 MHz) offers a sufficient guard band to avoid the effect of C-band ACI. Hence, all PCIs on BS-3 exhibit similar throughput, RSRP and RSRQ behavior. The obtained throughput levels at BS-3, including the peak throughput on PCI 88 (3560 MHz), is much lower compared to BS-1 and BS-2 due to the lower tower height. 

As shown in Fig. \ref{Fig:CBRSdeployment}, BS-3's PCI 96 faces west. Due to the short distance between BS-3 and BS-4, 1.6 km, and lower tower height of BS-3, this poses a potential CCI threat to PCI 187 in BS-4 operating on the same frequency.

\begin{figure}
	\centering
        \begin{subfigure}[]
		{\includegraphics[width=7 cm,height= 5 cm]{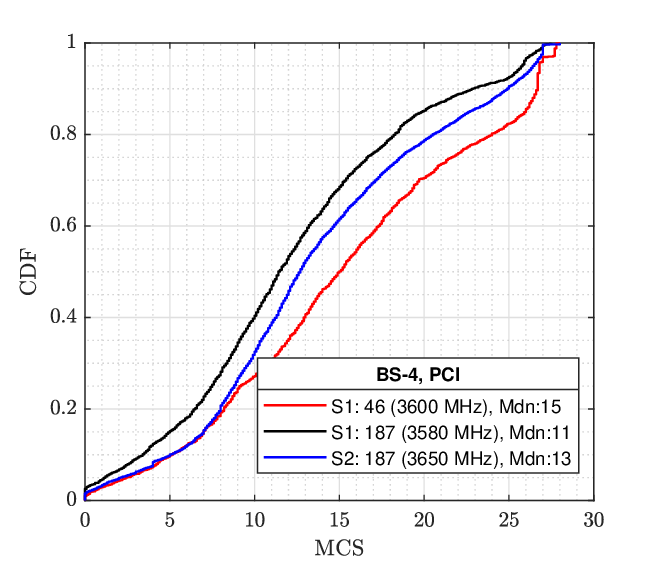}\label{Fig:Nav_MCS}}
	\end{subfigure} 
	\begin{subfigure}[]
		{\includegraphics[width=7 cm,height= 5 cm]{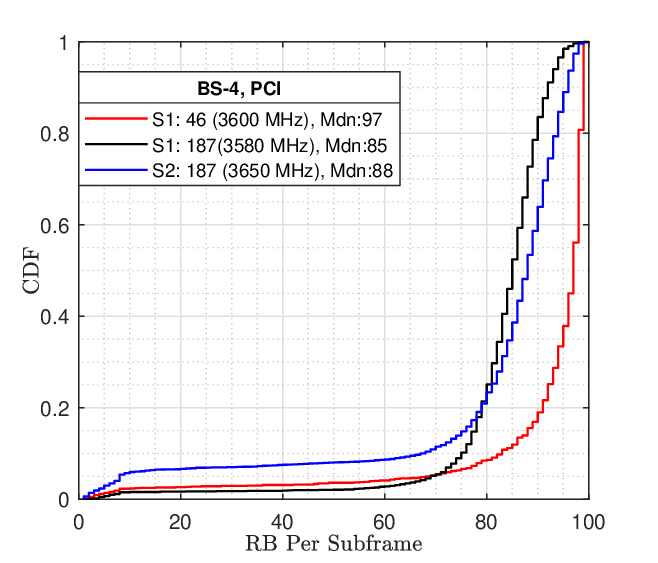}\label{Fig:Nav_RB_187}}
	\end{subfigure} 
	\caption{ MCS and RBs per subframe for PCIs at BS-4. } 
	\label{Fig:Navarre_2_RB}
\end{figure}

%
%
\subsection{Avoiding CCI by selecting an appropriate frequency}

MC-2 on PCI 187 evaluated the impact of CCI within the CBRS deployment itself and aims to improve system performance via a suitable frequency allocation. 

\textit{Performance of BS-4:} PCI 187 from BS-4, operating on 3580 MHz (S1) shows the worst throughput, RSRP and RSRQ due to CCI coming from PCI 96 of BS-3, as shown in Fig. \ref{Fig:Navarre_1}, while PCI 46 from the same BS does not experience CCI and exhibits nearly the same performance as the PCIs on BS-3 as it is the only PCI operating on 3600 MHz in the region of BS-3 and BS-4. 

To alleviate the effect of CCI on PCI 187, we proposed changing the frequency from 3580 MHz (S1) to 3650 MHz (S2) based on our study of frequency allocations and measurements of signal strengths in the vicinity of the CBRS deployment. The frequency 3650 MHz was not used by any of the BSs deployed by the school district, as seen from Table \ref{tab:Freq_vs_PCI}. As illustrated in Fig.~\ref{Fig:vrz_cbrs_map}, Verizon has utilized the frequency 3650 MHz in the region of BSs, but not in the direction of PCI 187. 

Fig.~\ref{Fig:Nav_RSRP} and Fig.~\ref{Fig:Nav_RSRQ} show that changing the frequency of PCI 187 to 3650 MHz resulted in higher RSRP and RSRQ levels compared to the original frequency of 3580 MHz, and a similar performance to PCI 46, which is free of CCI. It is important to highlight that PCI 46 maintained the same performance after the frequency change on PCI 187, as they do not operate on the same frequency.

As compared with S1, the peak throughput of S2 increased by around 20 Mbps, from 100 Mbps to 120 Mbps, while the median throughput remained the same as shown in Fig.~\ref{Fig:Nav_through}. This can be explained as follows: throughput is determined by the MCS and number of resource blocks (RBs) allocated per subframe, as shown in Fig.~\ref{Fig:Navarre_2_RB}.
In Fig. \ref{Fig:Nav_MCS}, the median value of MCS for S2 increased by 2 indicating that the frequency 3650 MHz is exposed to less CCI compared to the frequency 3580 MHz, and the number of RBs also improved slightly, but not enough to deliver a significant throughput increase. We see that PCI 46 for example has a higher number of RBs/subframe leading to consistently higher throughput. We speculate that the number of RBs/subframe allocated to PCI 187 were lower than PCI 46 due to proprietary network optimization algorithms, and hence, even though the signal metrics, RSRP, RSRQ and MCS all improved with the change in frequency, the resulting median throughput remained unchanged, though the maximum throughput did improve.

\begin{figure}
	\centering
	\includegraphics[width=7 cm,height= 5 cm]{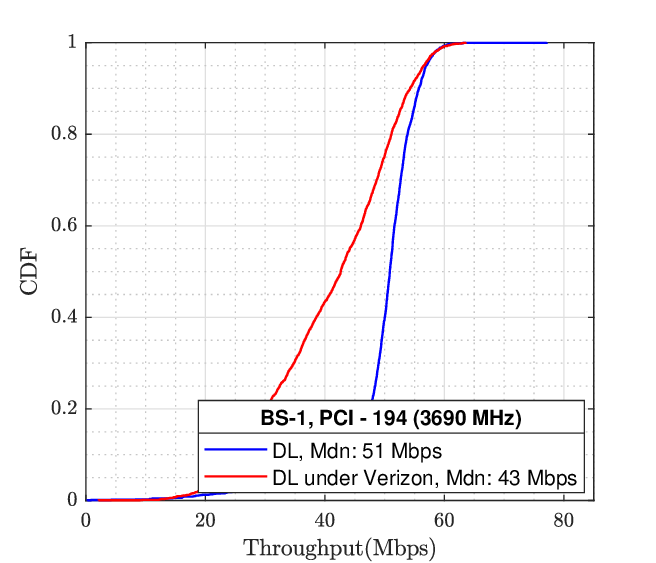}
	\caption{The impact of C-Band on PCI 194 (3690 MHz) at BS-1.}
	\label{Fig:Cband_Interference}
\end{figure}

%
%
\subsection{Impact of ACI from C-band on CBRS} 

MC-3 on PCI 194 evaluates the effect of ACI caused by C-band on CBRS. We performed stationary measurements in the location shown in Fig.~\ref{Fig:loc-test3} where the strongest RSRP was measured for PCI 194 on 3690 MHz. 

Fig.~\ref{Fig:Cband_Interference} shows the DL throughput performance when connected to PCI 194 in BS-1, in the absence and presence of a Verizon C-band UE. First, we measured DL throughput on the UE connected to CBRS only, followed by simultaneous DL transmissions to two UEs, one connected to CBRS and the other connected to C-band. We observed around 16\% throughput degradation on the CBRS UE due to ACI from C-band, when both devices were simultaneously connected. The CBRS UE achieved a peak throughput of approximately 80 Mbps when there was no transmission from the C-band user. However, its maximum throughput was limited to around 60 Mbps in the presence of the C-band UE. The absence of guard bands between CBRS and C-band, the transmit power difference and lower tower height of BS-1 compared to the Verizon C-band, all contribute to the reduced throughput performance on CBRS UE due to adjacent channel C-band usage. 

\begin{figure}
	\centering
	\begin{subfigure}[]
		{\includegraphics[width=7 cm,height= 4 cm]{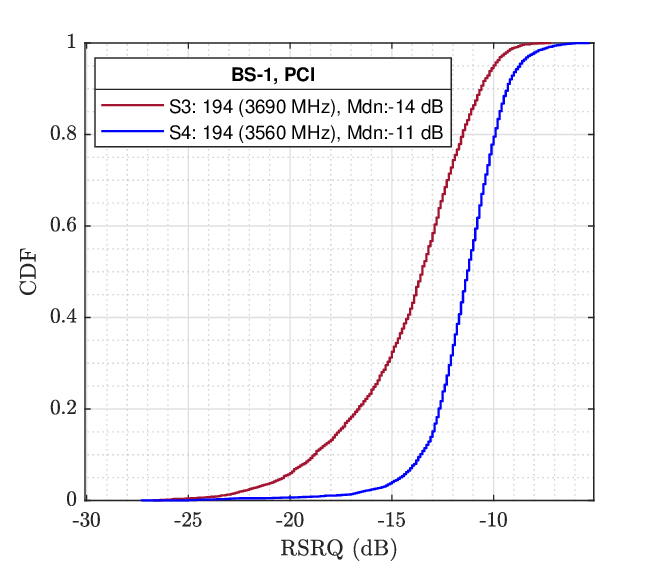}\label{Fig:PCI194_FreqChangeRSRQ}}
	\end{subfigure} 
 \begin{subfigure}[]
		{\includegraphics[width=7 cm,height= 4 cm]{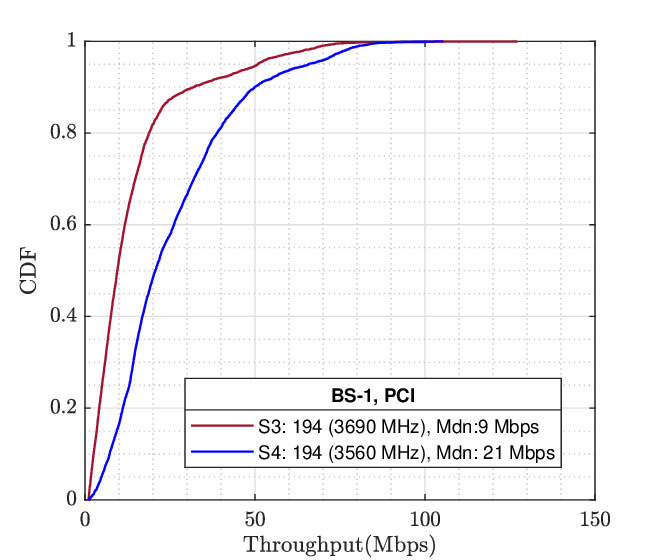}\label{Fig:PCI194_FreqChangeThroughput}}
	\end{subfigure} 
	\caption{RSRQ and throughput for PCI 194 i) Scenario 3 (S3): 3690 MHz, and ii) Scenario 4 (S4): 3560 MHz.}
	\label{Fig:PCI194_FreqChange}
\end{figure}

Fig.~\ref{Fig:PCI194_FreqChange} assesses the performance improvement of PCI 194 after a frequency change from 3690 MHz (S3) to 3560 MHz (S4): this change was made in response to our measurements that indicated significant adjacent channel interference from C-band deployments in the vicinity. We performed a driving measurement campaign in the coverage area of PCI 194 after the frequency change to evaluate the influence of appropriate frequency allocations on mitigating ACI from C-band in CBRS band.
Fig.~\ref{Fig:PCI194_FreqChangeRSRQ} shows that changing to S4 resulted in higher median RSRQ levels compared to the original frequency of 3690 MHz: an increase from -14 dB to -11 dB. Moreover, S4 now exhibits comparable RSRQ performance to the neighboring PCI 150 of BS1, which is free of ACI. As compared with S3, the median throughput of S4 increased by around 12 Mbps, from 9 Mbps to 21 Mbps, while the peak throughput remained the same as shown in Fig.~\ref{Fig:PCI194_FreqChangeThroughput}.

A similar scenario will arise between 3.45 GHz and CBRS once 5G deployments in the 3.45 - 3.55 GHz band increase, indicating further performance degradation on CBSDs operating at the lower edge of the CBRS band. Hence CBRS deployments need to know whether 5G is being deployed in upper and/or lower adjacent bands and not use the band-edge CBRS channels which will be impacted the most by ACI. Since the SAS does not provide this information, this intelligence needs to be available at the deployment site using measurements tools such as the ones used in this study.

\subsection{CBRS band use by Verizon using LTE-CA}

\begin{figure}
	\centering
	\begin{subfigure}[CBRS and LTE-CA throughput.]
		{\includegraphics[width=7 cm,height= 4 cm]{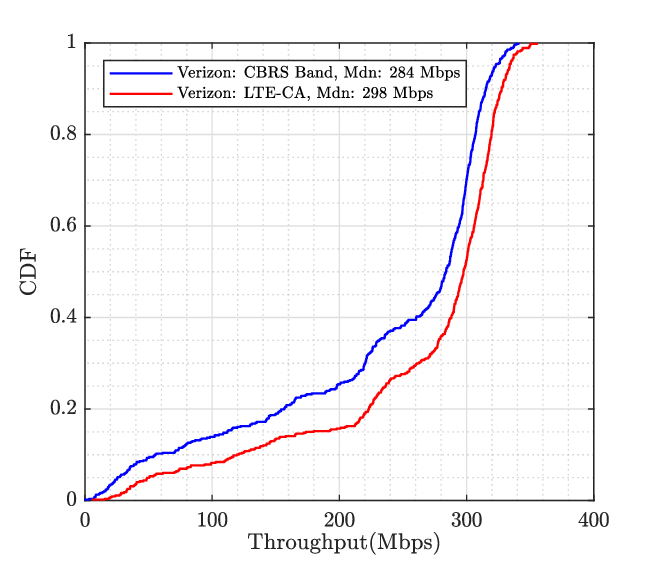}\label{Fig:MC3_Ver_Agg}}
	\end{subfigure} 
 \begin{subfigure}[Ratio of CBRS to total LTE-CA throughput.]
		{\includegraphics[width=7 cm,height= 4 cm]{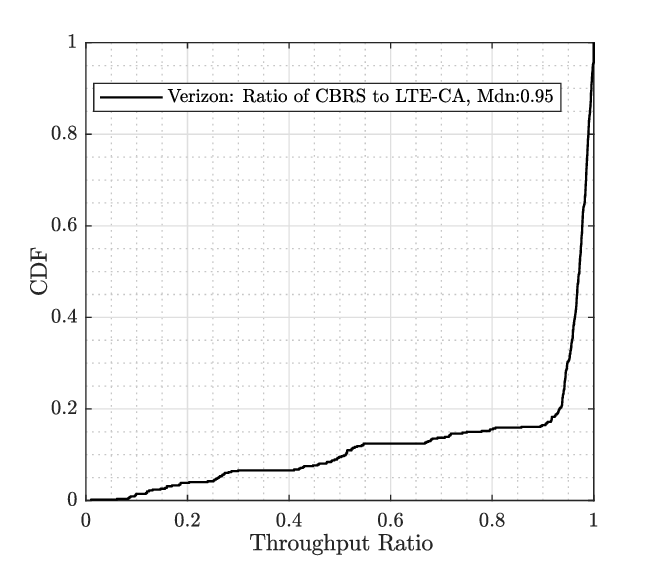}\label{Fig:MC3_Ratio}}
	\end{subfigure} 
	\caption{Verizon throughput with LTE-CA using CBRS.} 
	\label{Fig:LTE_CA_Discussion}
\end{figure}

\begin{figure}
	\centering
	\includegraphics[width=7 cm,height= 4 cm]{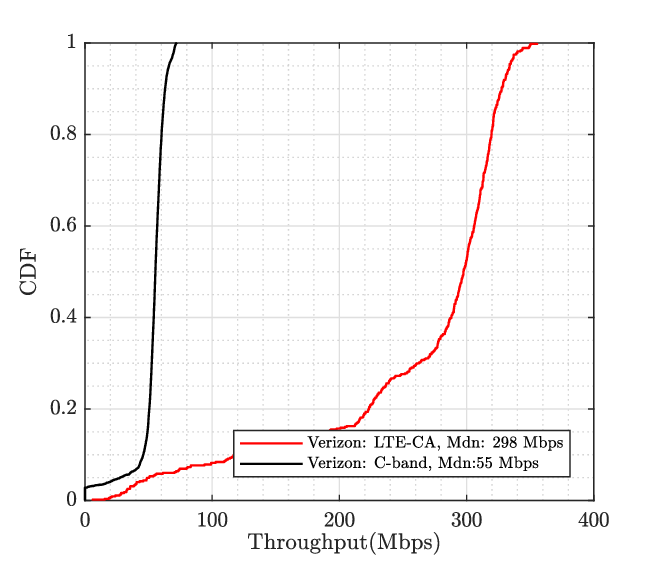}
	\caption{Verizon LTE-CA and C-band throughput.}
	\label{Fig:Cband_Tput}
\end{figure}

During our experiments for MC-3, we observed that the Verizon BS also transmits on the CBRS band using LTE with carrier aggregation (LTE-CA), aggregating up to five 20 MHz CBRS channels with four channels being aggregated most often (80\%). Along with a 20 MHz primary LTE channel, this allows up to 120 MHz of bandwidth for use when high capacity is required: this is significantly higher than the C-band channel bandwidth of 60 MHz that Verizon has exclusive license to, but at zero cost compared to the billions of dollars spent for exclusive licenses.
Table \ref{tab:features_by_Verizon} details how the CBRS band (also called Band 48), is utilized by Verizon in the vicinity of the South Bend CBRS deployment. We see that Verizon CBSDs are deployed on all available CBRS frequencies, creating potential CCI for other CBRS deployments. Additionally, we observed that when LTE-CA uses CBRS channels, the total throughput as well as the proportion of throughput carried over CBRS are both very high, as shown in Fig.~\ref{Fig:LTE_CA_Discussion} and overall throughput of 4G using LTE-CA was significantly higher than 5G using C-band at the same location as shown in Fig.~\ref{Fig:Cband_Tput}. Thus, even as operators roll out 5G using their new, licensed, spectrum, CBRS remains extremely competitive when additional capacity is needed.

\begin{table}
	\caption{Frequencies used by Verizon in the vicinity of the South Bend CBRS deployment.}
	\centering
	\small	
\renewcommand{\arraystretch}{1}
	\begin{tabular}{|C{1.5 cm}|C{2.5cm}||C{1.5cm}|C{1.5cm}|} 
 \hline
 \textbf{Band} &  \textbf{Freq. (MHz)} &  \textbf{Band} &  \textbf{Freq. (MHz)} \\
  \hline \hline  CBRS, Band 48 & 3560, 3570, 3580, 3590, 3600, 3610, 3610, 3620, 3630, 3640, 3650, 3660, 3670, 3680 & C-band, \newline Bands n77/n78 & 3730, 3809 \\
\hline
	\end{tabular}
	\label{tab:features_by_Verizon} 
\end{table}

%
%
\section{Conclusion}


The extensive measurements and analyses presented in this study conclusively demonstrate that secondary coexistence among GAA CBSDs, even when they belong to the same CBRS network, can be a limiting factor for optimal performance. In the deployment we studied, all CBRS channels were available according to the Google SAS, however the South Bend CBRS deployment used three 20 MHz channels (3580 MHz, 3670 MHz and 3690 MHz) more often than the others. Further, the deployment did not take into account the emergence of Verizon CBRS using GAA mode in the vicinity, along with adjacent channel C-band, both of which further impacted the performance. In order to demonstrate the impact of appropriate frequency allocation, we worked with the CBRS provider to change the frequency of one CBSD and demonstrated improved performance of signal quality metrics. However, this change took a while to implement since the SAS had to authorize the new channel. Thus, it is clear that CBRS deployments need to be able to dynamically change their operating channel based on measurements in the field: such dynamic behavior is not possible today since all channel allocations must be through the SAS provider. Further, we demonstrated that even with 4G, when multiple CBRS channels were aggregated, the throughput was higher than that obtained with 5G using C-band, thus demonstrating the usefulness of CBRS to both large mobile operators as well as small private network providers such as the South Bend school district. Future work will consider how channel choice by CBRS can be made more dynamic and other ways to better manage secondary coexistence among GAA users.

%
%
\section*{Acknowledgment}
This work was made possible by the cooperation of the South Bend Community School Corporation and the CBRS network provider.
\bibliographystyle{IEEEtran}
\bibliography{References}

\begin{thebibliography}{10}
\providecommand{\url}[1]{#1}
\csname url@samestyle\endcsname
\providecommand{\newblock}{\relax}
\providecommand{\bibinfo}[2]{#2}
\providecommand{\BIBentrySTDinterwordspacing}{\spaceskip=0pt\relax}
\providecommand{\BIBentryALTinterwordstretchfactor}{4}
\providecommand{\BIBentryALTinterwordspacing}{\spaceskip=\fontdimen2\font plus
\BIBentryALTinterwordstretchfactor\fontdimen3\font minus \fontdimen4\font\relax}
\providecommand{\BIBforeignlanguage}[2]{{%
\expandafter\ifx\csname l@#1\endcsname\relax
\typeout{** WARNING: IEEEtran.bst: No hyphenation pattern has been}%
\typeout{** loaded for the language `#1'. Using the pattern for}%
\typeout{** the default language instead.}%
\else
\language=\csname l@#1\endcsname
\fi
#2}}
\providecommand{\BIBdecl}{\relax}
\BIBdecl

\bibitem{shah20185g}
S.~A.~A. Shah, E.~Ahmed, M.~Imran, and S.~Zeadally, ``{5G for vehicular communications},'' \emph{IEEE Communications Magazine}, vol.~56, no.~1, pp. 111--117, 2018.

\bibitem{ghosh2019competition}
A.~Ghosh and R.~Berry, ``{Competition with three-tier spectrum access and spectrum monitoring},'' in \emph{Proceedings of the 20th ACM International Symposium on Mobile Ad Hoc Networking and Computing}, 2019, pp. 241--250.

\bibitem{exoplanetwebsite}
\BIBentryALTinterwordspacing
{The Wireless Innovation Forum}. (2023) {CBRS Operational Security. WINNF-TS-0071.} [Online]. Available: \url{https://cbrs.wirelessinnovation.org/release-1-standards-specifications}
\BIBentrySTDinterwordspacing

\bibitem{agarwal2022survey}
P.~Agarwal, M.~Manekiya, T.~Ahmad, A.~Yadav, A.~Kumar, M.~Donelli, and S.~T. Mishra, ``{A survey on citizens broadband radio service (CBRS)},'' \emph{Electronics}, vol.~11, no.~23, p. 3985, 2022.

\bibitem{ross2019annual}
W.~L. Ross and S.~D.~W. Kinkoph, ``{Annual report on the status of spectrum repurposing},'' \emph{Proceedings of the 2019, US Department of Commerce}, 2019.

\bibitem{dogan2023evaluating}
S.~Dogan-Tusha, M.~I. Rochman, A.~Tusha, H.~Nasiri, J.~Helzerman, and M.~Ghosh, ``{Evaluating the interference potential in 6 GHz: An extensive measurement campaign of a dense indoor Wi-Fi 6E network},'' in \emph{Proceedings of the 17th ACM workshop on wireless network testbeds, experimental evaluation \& characterization}, 2023, pp. 56--63.

\bibitem{paolini2019cbrs}
M.~Paolini and S.~Fili, ``{CBRS: Should the enterprise and venue owners care?}'' \emph{Senza Fili}, 2019.

\bibitem{kim2015design}
C.~W. Kim, J.~Ryoo, and M.~M. Buddhikot, ``{Design and implementation of an end-to-end architecture for 3.5 GHz shared spectrum},'' in \emph{2015 IEEE International Symposium on Dynamic Spectrum Access Networks (DySPAN)}.\hskip 1em plus 0.5em minus 0.4em\relax IEEE, 2015, pp. 23--34.

\bibitem{maeng2023sdr}
S.~J. Maeng, O.~Ozdemir, {\.I}.~G{\"u}ven{\c{c}}, M.~L. Sichitiu, M.~Mushi, R.~Dutta, and M.~Ghosh, ``{SDR-based 5G NR C-band i/q monitoring and surveillance in urban area using a helikite},'' in \emph{2023 IEEE International Conference on Industrial Technology (ICIT)}.\hskip 1em plus 0.5em minus 0.4em\relax IEEE, 2023, pp. 1--6.

\bibitem{verizon}
\BIBentryALTinterwordspacing
{Verizon}. (2023) {The map of approximate outdoor coverage of 5G Ultra Wideband for Verizon in USA}. [Online]. Available: \url{https://www.verizon.com/coverage-map/}
\BIBentrySTDinterwordspacing

\bibitem{rochman2023measurement}
M.~I. Rochman, V.~Sathya, B.~Payne, M.~Yavuz, and M.~Ghosh, ``A measurement study of the impact of adjacent channel interference between c-band and cbrs,'' in \emph{2023 IEEE 34th Annual International Symposium on Personal, Indoor and Mobile Radio Communications (PIMRC)}.\hskip 1em plus 0.5em minus 0.4em\relax IEEE, 2023, pp. 1--7.

\bibitem{9261954}
V.~Sathya, M.~I. Rochman, and M.~Ghosh, ``{Measurement-based coexistence studies of LAA \& Wi-Fi deployments in Chicago},'' \emph{IEEE Wireless Communications}, vol.~28, no.~1, pp. 136--143, 2021.

\bibitem{QualiPocref}
\BIBentryALTinterwordspacing
{Rohde \& Schwarz}. (2023) {QualiPoc: Handheld mobile network testing and trouble shooting}. [Online]. Available: \url{https://www.rohde-schwarz.com/us/products/test-and-measurement/network-data-collection/qualipoc-android\_63493-55430.html}
\BIBentrySTDinterwordspacing

\bibitem{prismref}
\BIBentryALTinterwordspacing
{EPiQ Solutions}. (2023) {PRiSM: Handheld network scanner and spectrum analyzer}. [Online]. Available: \url{https://epiqsolutions.com/rf-sensing/prism/}
\BIBentrySTDinterwordspacing

\end{thebibliography}

\end{document}